\documentclass[10pt, conference]{IEEEtran}
\usepackage{epsfig,color,amsmath,cite}
\usepackage{amsthm}
\usepackage{amsmath}    
\IEEEoverridecommandlockouts
\usepackage{bm}
\usepackage{epstopdf}
\usepackage{amssymb}
\usepackage{url}
\usepackage{enumitem}
\usepackage{multirow}
\usepackage{hhline}
\usepackage{booktabs}
\usepackage{threeparttable}
\usepackage{caption}
\usepackage[linesnumbered,lined,boxed,commentsnumbered,ruled,longend]{algorithm2e}
\usepackage{comment}
\setlist[itemize]{leftmargin=*}

\DeclareMathOperator*{\minimize}{minimize}

\DeclareMathOperator*{\find}{find}
\DeclareMathOperator*{\subjectto}{subject\ to}

\makeatother
\DeclareMathAlphabet\mathbfcal{OMS}{cmsy}{b}{n}



\usepackage{stackengine}

\newcommand{\bmat}[1]{\begin{bmatrix} #1 \end{bmatrix}}

\newcommand{\m}{\boldsymbol}
\allowdisplaybreaks[4]
\usepackage[colorlinks = true,
linkcolor = blue,
urlcolor  = blue,
citecolor = blue,
anchorcolor = blue]{hyperref}

\newcommand{\mbb}[1]{\mathbb{#1}}

\usepackage[framemethod=TikZ]{mdframed}
\mdfdefinestyle{MyFrame}{%
	linecolor=black,
	outerlinewidth=1.25pt,
	roundcorner=1.25pt,
	innerrightmargin=5pt,
	innerleftmargin=5pt,}
	
\usepackage[noabbrev]{cleveref}

\usepackage{mathtools}

\DeclarePairedDelimiter\abs{\lvert}{\rvert}%
\DeclarePairedDelimiter\norm{\lVert}{\rVert}%

\makeatletter
\let\oldabs\abs
\def\abs{\@ifstar{\oldabs}{\oldabs*}}
\let\oldnorm\norm
\def\norm{\@ifstar{\oldnorm}{\oldnorm*}}
\makeatother

\usepackage[english]{babel}
\usepackage[utf8]{inputenc}
\usepackage[super]{nth}

\usepackage{graphicx}
\usepackage{float}
\usepackage[caption = false]{subfig}

\newcommand*{\newqed}{\null\nobreak\hfill\ensuremath{\blacksquare}}%

\title{\vspace{0.4cm} \LARGE \textbf{Feedback Stabilization and Output Tracking for Discrete-Time Lipschitz Nonlinear Systems via Iterative Convex Approximations}}

\author{Sebastian A. Nugroh$\text{o}^{\dagger}$, Suyash C. Vishno$\text{i}^{\ddagger}$, Ahmad F. Tah$\text{a}^{\dagger}$, and Christian G. Claude$\text{l}^{\ddagger}$ \vspace{-0.5cm}
	\thanks{
    	$^{\dagger}$Department of Electrical and Computer Engineering, The University of Texas at San Antonio, 1 UTSA Circle, San Antonio, TX 78249.
		$^\ddagger$Department of Civil, Architectural, and Environmental Engineering, The University of Texas at Austin, 301 E. Dean Keeton St. Stop C1700, Austin, TX 78712.
		Emails: sebastian.nugroho@my.utsa.edu, scvishnoi@utexas.edu, ahmad.taha@utsa.edu, christian.claudel@utexas.edu, taposh.banerjee@utsa.edu.
		This work was partially supported by the National Science Foundation under Grants 1636154, 1728629, 1917164, 1917056, and 2013786.}
}

\begin{document}

\maketitle

\setlength{\abovedisplayskip}{3.1pt}
\setlength{\belowdisplayskip}{3.1pt}
\setlength{\abovedisplayshortskip}{3.1pt}
\setlength{\belowdisplayshortskip}{3.1pt}

\newdimen\origiwspc%
\newdimen\origiwstr%
\origiwspc=\fontdimen2\font
\origiwstr=\fontdimen3\font

\fontdimen2\font=0.63ex

\begin{abstract}
The stabilization of unstable nonlinear systems and tracking control are challenging engineering problems due to the encompassed nonlinearities in dynamic systems and their scale. In the past decades, numerous observer-based control designs for dynamic systems in which the nonlinearity belongs to Lipschitz functions have been proposed. However, most of them only focus on output feedback and consequently, state feedback design remains less developed. To that end, this paper is dedicated to the problem of full-state feedback controller design for discrete-time Lipschitz nonlinear systems. In addition, we present a simple iterative method for improving the convergence of the closed-loop performance. It is later demonstrated that our approach can be conveniently extended and utilized for output tracking.
\end{abstract}

\section{Introduction}
The problems of tracking and stabilization for nonlinear systems have been widely studied in the literature over the past decades. The majority of the research focuses on asymptotic stabilization of the system with the help of output feedback or observer-based controllers designed for a class of Lipschitz nonlinear systems. Some of the noticeable works in this area for discrete-time systems include \cite{grandvallet2012observer, yu2020observer ,ibrir2005observer,abbaszadeh2016static,ho2003robust,NguyenCuongM2019Roao}. The general advantages of output feedback over state feedback controllers has led to a lack of exploration of state feedback control for discrete-time nonlinear systems. However, it cannot be denied that the technique is still efficient under favorable settings, that is, under the availability of full state information. Therefore, in this paper, we propose the design of a full-state feedback controller for stabilization of the said class of systems. In doing so, we consider that the nonlinearity is a function of both the state and the inputs rather than only the state as is the case in some studies, for instance, \cite{grandvallet2012observer,yu2020observer,ibrir2005observer}. Besides, the proposed controller is easily extendable for output tracking along with stabilization.

\indent Some developments have been made in the literature for this particular research direction.
For instance, the authors in \cite{StipanovicD.M2001Rsas} present an asymptotic stabilization approach using state feedback for discrete-time Lipschitz nonlinear systems. 
A stabilization and tracking-based control method for continuous-time Lipschitz nonlinear systems is proposed in \cite{rehan2011stabilization}.
In \cite{LashariMuhammadAfzal2015Satc}, inspired by \cite{rehan2011stabilization}, the authors propose a state feedback controller for stabilization and tracking control of discrete-time Lipschitz nonlinear systems. It is worth noting that most of these studies assume that the nonlinearity is only a function of the state. 
Moreover, their method requires solving bilinear matrix inequalities (BMIs) which is not possible using the readily available convex programming solvers. While the controller design in the present article does make use of BMIs, we propose a simple method to find a potential feasible solution by solving a series of linear matrix inequalities (LMIs) instead of a BMI.

Some other studies that explore the idea of state feedback control in Lipschitz nonlinear systems are \cite{YadegarMeysam2018Otcd,PagillaP.R2004Caod,ekramian2017observer} which do so in a continuous-time setting. While the controllers proposed in these works cannot be directly implemented for the discrete-time case, the methodology used for design, specifically in \cite{ekramian2017observer}, has inspired the design of the controller presented in this article. The contributions of this work are twofold: \textit{(a)} the development of a full-state feedback control approach for achieving exponential stability for discrete-time Lipschitz nonlinear systems where the nonlinearity is dependent on both the states and the inputs. In addition, the proposed framework can also be easily used for the purpose of output tracking control; \textit{(b)} an iterative method is proposed for solving the stabilization problem as well as maximizing the convergence rate of the controller by means of successive convex approximation (SCA), which basically over-approximates the nonconvex terms in the feasibility problem with a series of LMIs.
The remainder of the paper is structured as follows. In Section \ref{sec:formulation}, we present the state-space representation for the class of Lipschitz nonlinear systems considered in this paper. While the full-state feedback controller design for stabilization and output regulation is given in Section \ref{sec:controllerdesign}, the algorithmic procedure for computing the controller gain and improving the closed-loop convergence rate are presented in Section \ref{sec:solution_algorithm}. Next, numerical examples are presented in Section \ref{sec:simulationresults} and finally Section \ref{sec:conclusion} concludes the paper.



\section{ Preliminaries and Problem Formulation\vspace{-0.0cm}}\label{sec:formulation}
Consider the following nonlinear dynamical system
\begin{subequations}\label{e:system}
	\begin{align}
	\m x[k+1]&=\m A \m x[k]+\m G\m f(\m x[k],\m u[k])+\m B\m u[k], \\
	\m y[k]&=\m C\m x[k],
\end{align}
\end{subequations}
where in \eqref{e:system}, $\m x\in \mathbb{R}^n$ is the state vector, $\m u\in\mathbb{R}^m$ is the control input vector, and $\m y\in\mathbb{R}^p$ is the output vector. Matrices $\m A\in\mathbb{R}^{n\times n}, \m G\in\mathbb{R}^{n\times g}, \m B\in\mathbb{R}^{n \times m}$, and $\m C\in\mathbb{R}^{p\times n}$ are assumed to be known and constant. It is assumed that 
$\m f:\mathbb{R}^n\times\mathbb{R}^m\rightarrow\mathbb{R}^g$ is a Lipschitz function with respect to state $\m x$ and control input $\m u$. That is, there exist non-negative constants $\gamma_x \geq 0$ and $\gamma_u\geq 0$ such that for all $\m x_1,\m x_2\in \mathbb{R}^n$ and $\m u_1,\m u_2\in \mathbb{R}^m$
\begin{subequations}\label{e:lipschitz}
\begin{align}
    \lvert\lvert \m f(\m x_1,\m u) - \m f(\m x_2,\m u)\rvert\rvert_2\le\gamma_x\lvert\lvert \m x_1 - \m x_2\rvert\rvert_2\\
    \lvert\lvert \m f(\m x,\m u_1) - \m f(\m x,\m u_2)\rvert\rvert_2\le\gamma_u\lvert\lvert \m u_1 - \m u_2\rvert\rvert_2
\end{align}
\end{subequations}
In this paper, we design two types of controllers for discrete-time nonlinear systems like the one described in \eqref{e:system}, one that stabilizes the output to some arbitrary equilibria, and another that not only stabilizes the output but also makes it to converge to a reference point. In the ensuing section, a state feedback controller is designed for the system defined in \eqref{e:system}. The following lemma is instrumental for the development of our approach. 



\noindent \textbf{Lemma 1.} (\hspace{-0.01cm}\cite{Lee2016}). \textit{Let $\mathcal{ S} : \mbb{R}^n\rightarrow \mbb{R}^q$ be a linear mapping such that $\mathcal{ S}(\m x):= \sum_{i = 1}^{n} x_i \m S_i$ where $\m S_i\in\mbb{S}^q$ for $i\in\{1,2,\hdots,n\}$. If it holds that $\mathcal{ S}(\m x) \succ 0,\;\; \mathcal{ S}(\m y)\succ 0$ for $\m x,\m y\in\mbb{R}^n$, then}
\begin{align}
	-\mathcal{ S}(\m y)^{-1}\preceq -2\mathcal{ S}(\m x)^{-1} + \mathcal{ S}(\m x)^{-1}\mathcal{ S}(\m y)\mathcal{ S}(\m x)^{-1}.
\end{align}



\vspace{-0.2cm}
\section{Full State Feedback Controller Synthesis}\label{sec:controllerdesign}
In this section, we propose a state feedback controller for the system given in \eqref{e:system}. 
We first consider a regulation problem with a state feedback controller. The state feedback control law is designed as follows
\begin{align}
    \m u[k]=-\m K\m x[k], \label{e:input}
\end{align}
where $\m K\in\mathbb{R}^{m\times n}$ is the constant matrix gain chosen such that the nonlinear discrete-time system \eqref{e:system} is asymptotically stable. Before we present the proposed approach, the following assumption is considered throughout the paper.

\noindent \textbf{Assumption 1.} \textit{There exists an equilibrium point $\m x_{eq}\in\mbb{R}^n$ for \eqref{e:system} with control law \eqref{e:input} such that}
\begin{align}
\m x_{eq} &= \m A \m x_{eq}+\m G\m f(\m x_{eq},\m u_{eq})+\m B\m u_{eq}, \label{e:equilibrium_dynamics}
\end{align}
\textit{where $\m u_{eq}=-\m K\m x_{eq}$ is the input vector at equilibrium.} 

Let us define $\m e\in\mbb{R}^n$ as $\m e := \m x - \m x_{eq}$. That is, $\m e$ represents the deviation of the state $\m x$ from its equilibrium $\m x_{eq}$. From \eqref{e:system} and \eqref{e:equilibrium_dynamics} with control law as given in \eqref{e:input}, we obtain the following closed-loop error dynamics 
\begin{align}
	\m e[k+1] &= (\m A - \m B \m K)\m e[k] + \m G \Delta \m f[k],\label{eq:closed_loop_error_dyn}
\end{align}
where $	\Delta \m f[k] := \m f(\m x[k],\m u[k]) - \m f(\m x_{eq},\m u_{eq}).$
Now, our objective is to find a computationally amenable way to compute $\m K$ such that the system \eqref{eq:closed_loop_error_dyn} is (at least) asymptotically stable. To that end, in what follows we present a new sufficient condition to achieve exponential stability for \eqref{eq:closed_loop_error_dyn}.

\noindent \textbf{Theorem 1.} \textit{Consider nonlinear discrete-time system \eqref{eq:closed_loop_error_dyn} with control law presented in \eqref{e:input}. The system in \eqref{eq:closed_loop_error_dyn} is exponentially stable if there exist $\m Q\in\mathbb{S}^n,\;\m K\in\mathbb{R}^{m\times n},\;\epsilon,\kappa,\alpha\in\mathbb{R}$ which solve the following feasibility problem}
\begin{subequations}\label{eq:SDP_base}
	\begin{align}
&\bmat{(\alpha-1)\m Q+\epsilon(\gamma_x+\gamma_u\kappa)^2\m I & * & *\\
	\m O & -\epsilon \m I &*\\
	\m A-\m B\m K & \m G & -\m Q^{-1}}\prec 0 \label{eq:SDP_base_1}\\ 
&\bmat{-\kappa \m I & \m K \\ \m K^\top & -\kappa \m I} \preceq 0 \label{eq:SDP_base_3}\\
&\m Q \succ 0, \epsilon\ge 0, \kappa>0, 0 < \alpha < 1. \label{eq:SDP_base_2}
\end{align}
\end{subequations}
\noindent \textit{Proof.} Consider the nonlinear discrete-time system \eqref{eq:closed_loop_error_dyn}. Note that the term $\Delta \m f(\cdot)$ is equal to
\begin{align}\label{e:Lipschitz_Continuity_Application}
    &\lvert\lvert\Delta \m f\rvert\rvert_2=\lvert\lvert \m f(\m x,\m u)-\m f(\m x_{eq},\m u_{eq})\rvert\rvert_2 \nonumber\\ 
    &\quad=\lvert\lvert (\m f(\m x,\m u)-\m f(\m x_{eq},\m u))+(\m f(\m x_{eq},\m u)-\m f(\m x_{eq},\m u_{eq}))\rvert\rvert_2 \nonumber\\
    &\quad\le\lvert\lvert \m f(\m x,\m u)-f(\m x_{eq},\m u)\rvert\rvert_2+\lvert\lvert \m f(\m x_{eq},\m u)-\m f(\m x_{eq},\m u_{eq})\rvert\rvert_2 \nonumber\\
    &\quad\le\gamma_x\lvert\lvert \m x-\m x_{eq}\rvert\rvert_2+\gamma_u\lvert\lvert \m u-\m u_{eq}\rvert\rvert_2 \nonumber\\
    &\quad=(\gamma_x+\gamma_u\lvert\lvert \m K\rvert\rvert_2)\lvert\lvert \m e\rvert\rvert_2
\end{align}
To design a stabilizing $\m K$, then consider a Lyapunov function candidate
$V : \mbb{R}^n \rightarrow \mbb{R}$ constructed as
\begin{align}
    V(\m e[k])=\m e[k]^\top \m Q\m e[k], Q\in \mathbb{S}^n.\label{eq:lyap_fun_cand}
\end{align}
for $Q\succ0$. First, we will show that the condition $\Delta V(\m e[k]) + \alpha V(\m e[k]) \leq 0$ for $0 < \alpha < 1$ leads to exponential stability.
Note that $\Delta V(\m e[k]) + \alpha V(\m e[k]) \leq 0$  for  $0 < \alpha < 1 $ implies
\begin{align}
	V(\m e[k+1]) \leq  (\alpha -1)V(\m e[k]).\, \label{eq:thm1_proof_1}
\end{align}
It can be shown by induction that the inequality \eqref{eq:thm1_proof_1} leads to
\begin{align}
	V(\m e[k]) \leq  (\alpha -1)^k V(\m e[0]).\, \label{eq:thm1_proof_2}
\end{align}
for $k \in \mbb{N}$. From \eqref{eq:lyap_fun_cand} and due to Rayleigh inequality, we have
\begin{align}
 	\lvert\lvert \m e[k]\rvert\rvert_2^2\leq \lambda_{\mathrm{min}}(\m Q)^{-1} V(\m e[k]). \label{eq:thm1_proof_3}
\end{align}
Now, from \eqref{eq:thm1_proof_2} and \eqref{eq:thm1_proof_3}, we obtain
\begin{align}
	\lvert\lvert \m e[k]\rvert\rvert_2^2 \leq (\alpha -1)^k \lambda_{\mathrm{min}}(\m Q)^{-1} V(\m e[0]). \label{eq:thm1_proof_4}
\end{align}
Since $V(\m e[0]) \leq \lambda_{\mathrm{max}}(\m Q)\lvert\lvert \m e[k]\rvert\rvert_2^2$, then from \eqref{eq:thm1_proof_4} we obtain
\begin{align}
	\lvert\lvert \m e[k]\rvert\rvert_2 \leq (\alpha -1)^{\frac{k}{2}} \sqrt{\frac{\lambda_{\mathrm{max}}(\m Q)}{\lambda_{\mathrm{min}}(\m Q)}} \lvert\lvert \m e[0]\rvert\rvert_2, \label{eq:thm1_proof_7}
\end{align}
showing that the system is exponentially stable. Second, it can be shown that the condition $\Delta V(\m e[k]) + \alpha V(\m e[k]) \leq 0$ for system \eqref{eq:closed_loop_error_dyn} is equivalent to 
\begin{align}
	\m \xi[k]^\top \m \Xi \m \xi[k] \leq 0, \label{eq:thm1_proof_5}
\end{align}
where $\m \xi[k] := \bmat{\m e[k]^\top & \Delta \m f[k]^\top}^\top$ for all $k \in \mbb{N}$ and 
\begin{align*}
\m \Xi :=	\bmat{(\m A - \m B \m K)^\top \m Q(\m A - \m B \m K)+ (\alpha-1)\m Q & * \\ \m G^\top \m Q(\m A - \m B \m K) & \m G^\top \m Q \m G}.
\end{align*}
From \eqref{e:lipschitz} and provided that $\lvert\lvert\m K\rvert\rvert_2\le \kappa^2$ for $\kappa > 0$, we have 
\begin{align}
	\m \xi[k]^\top \bmat{-(\gamma_x+\gamma_u\kappa)^2\m I & *\\ \m O & \m I}\m \xi[k] \leq 0. \label{eq:thm1_proof_6}
\end{align}
Applying the S-lemma \cite{yakubovich1997s} to \eqref{eq:thm1_proof_5} and \eqref{eq:thm1_proof_6} for $\epsilon \geq 0$, the following matrix inequality is obtained
\begin{align}
\bmat{\m	\Psi & * \\ \m G^\top \m Q(\m A - \m B \m K) & \m G^\top \m Q \m G - \epsilon \m I} \prec 0, \label{eq:thm1_proof_8}
\end{align}
where $\m\Psi$ is defined as
\begin{align*}
\m	\Psi &:= (\m A - \m B \m K)^\top \m Q(\m A - \m B \m K)+ (\alpha-1)\m Q \\ &\quad +\epsilon(\gamma_x+\gamma_u\kappa)^2\m I.
\end{align*}
By applying the Schur complement towards \eqref{eq:thm1_proof_8}, the matrix inequality \eqref{eq:SDP_base_1} is established. Finally, applying the Schur complement to the constraint $\lvert\lvert\m K\rvert\rvert_2\le \kappa^2$ for $\kappa > 0$ yields \eqref{eq:SDP_base_3}. This completes the proof. \newqed
\vspace{-0.0cm}

Note that the above deals with the stabilization of the discrete-time nonlinear dynamics to an unknown equilibrium. 
In some practical situations however, it is desirable to steer some of the states to follow some reference points. To achieve this, define an integrator as follows
\begin{align}
	\m e_y[k+1] = \m E \m e_y[k]. \label{eq:integrator_dynamics}
\end{align} 
In \eqref{eq:integrator_dynamics}, $\m e_y := \m r - \m C \m x$ where $\m r\in\mbb{R}^{p}$ represents the reference points and the matrix $\m E\in\mbb{R}^{p\times p}$ is a design parameter. It should be clear that $\m e_y$ denotes the error between the system's output and the reference signal. The incorporation of integrator \eqref{eq:integrator_dynamics} into the system \eqref{e:system} yields the following augmented dynamics 
\begin{align}
	\begin{split}
	\underbrace{\bmat{{{\m x}}[k+1] \\ {\m e}_y[k+1]}}_{\tilde{\m x}[k+1]} \hspace{-0.1cm}&=\hspace{-0.1cm}\underbrace{\bmat{\m A & \m O \\ \m E \m C &\m O}}_{\tilde{\m A}}\hspace{-0.05cm}\underbrace{\bmat{{{\m x}}[k] \\ {\m e}_y[k]}}_{\tilde{\m x}[k]}\hspace{-0.05cm}+\hspace{-0.05cm}\underbrace{\bmat{\m{B}\\ \m O}}_{\tilde{\m B}}\hspace{-0.05cm}{\m u}[k] \\
	&\quad+\hspace{-0.1cm}\underbrace{\bmat{\m G & \m O \\ \m O &\m E}}_{\tilde{\m G}}\hspace{-0.05cm}\underbrace{\bmat{{\m f}(\m x[k],\m u[k]) \\ -\m r[k]}}_{\tilde{\m f}}.
	\end{split}\label{eq:aug_dyn}
\end{align}
Notice that the augmented dynamics \eqref{eq:aug_dyn} share a similar form to that of \eqref{e:system}, which allows the design procedure described previously to be utilized in order to achieve stabilization and output regulation. It is worth noting that as the integrator \eqref{eq:integrator_dynamics} is comprised of linear dynamics, then thanks to Assumption 1, the solvability of \eqref{eq:SDP_base} with control structure \eqref{e:input} ensures output regulation to the reference points $\m r$.
Based on \eqref{eq:thm1_proof_7}, the following problem can be used to
maximize the stabilization's convergence rate 
\begin{subequations}\label{eq:SDP_opt_prob}
	\begin{align}
    (\mathbf{P1})\;\;&\minimize_{\m Q, \m K, \kappa, \epsilon,\alpha, t} \quad t \\
			&\subjectto \;\;\eqref{eq:SDP_base_1},\,\eqref{eq:SDP_base_3},\,\eqref{eq:SDP_base_2},\, \m Q \preceq t \m I,\, t > 0. 
	\end{align}
\end{subequations}

It is clear that either the feasibility problem posed in matrix inequalities \eqref{eq:SDP_base} or the optimization problem described in
 \eqref{eq:SDP_opt_prob} is difficult to solve since it contains nonconvex terms. In particular, the existence of matrix $\m Q$ and its inverse $\m Q^{-1}$ in 
\eqref{eq:SDP_base_1} together with the bilinear terms in $\alpha \m Q$ and  $\epsilon(\gamma_x+\gamma_u\kappa)^2$ constitute the problem's nonconvexity. 

\vspace{-0.2cm}
\section{Solution Approach via LMIs} \label{sec:solution_algorithm}
\vspace{-0.1cm}
The optimization problem $\mathbf{P1}$ and the feasibility problem described in \eqref{eq:SDP_base} require us to solve bilinear matrix inequalities (BMIs) which cannot be solved by means of a standard LMI solver like MOSEK \cite{Andersen2000}. On that regard, the following steps are proposed first to solve feasibility problem described in \eqref{eq:SDP_base}.\\
\textbf{Step 1:} For a fixed $\alpha\in(0,1)$ and predefined $\varrho < 0$, solve
\begin{subequations}\label{eq:step_1}
\begin{align}
    &\minimize_{\m X,\,\m Z,\,\nu}  \quad \nu \\
			&\subjectto \;\;\nonumber \\
			&\bmat{  (\alpha-1)\m X & \m A\m X-\m B\m Z\\
			\m X\m A^\top-\m Z^\top\m B^\top & -\m X} - \nu \m I\prec 0 \\
			&\m X \succ 0, \varrho \leq \nu < 0.
\end{align}
\end{subequations}
\noindent\textbf{Step 2:} For a fixed $\kappa > 0$, set $\m Q_0=\m X^{-1}$ and solve
\begin{subequations}\label{eq:step_2}
\begin{align}
    &\find \quad \epsilon \geq 0,\,\m K,\,0 <\alpha < 1\\
    & \subjectto \;\;\nonumber \\
    &\bmat{(\alpha-1)\m Q_0+\epsilon\gamma_{\kappa}^2\m I & 0 & \m A^\top-\m K^\top\m B^\top\\
    0 & -\epsilon \m I & \m G^\top\\
    \m A-\m B\m K & \m G & -\m Q_0^{-1}} \prec 0 \\
&\bmat{-\kappa \m I & \m K \\ \m K^\top & -\kappa \m I} \preceq 0,
\end{align}
\end{subequations}
where $\gamma_{\kappa} := \gamma_x+\gamma_u\kappa$. \\
The rationale behind Step 1 is motivated by the fact that the matrix $\m \Psi$ should be negative definite. Indeed, the matrix inequality $\m \Psi \prec 0$ is equivalent to the following condition
$$(\m A - \m B \m K)^\top \m Q(\m A - \m B \m K)+ (\alpha-1)\m Q  \prec -\epsilon(\gamma_x+\gamma_u\kappa)^2\m I.$$
Due to $\epsilon(\gamma_x+\gamma_u\kappa)^2$ being nonnegative and its value generally increases as the Lipschitz constants are increasing, the largest eigenvalue of $(\m A - \m B \m K)^\top \m Q(\m A - \m B \m K)+ (\alpha-1)\m Q$ has to be made sufficiently negative. Since $(\m A - \m B \m K)^\top \m Q(\m A - \m B \m K)+ (\alpha-1)\m Q \prec 0$ is equivalent to 
\begin{align}
	\bmat{  (\alpha-1)\m X & \m A\m X-\m B\m Z\\
		\m X\m A^\top-\m Z^\top\m B^\top & -\m X} \prec 0,\label{eq:step_1_LMI}
\end{align}
where $\m X = \m Q^{-1}$ and $\m Z = \m K\m Q^{-1}$. Thanks to the Schur complement lemma, then the problem described by \eqref{eq:step_1} is proposed, which bounds the left-hand side of \eqref{eq:step_1_LMI}. \\
\noindent \textbf{Remark 1.} \textit{If the problem described by \eqref{eq:step_2} is feasible for a fixed $\kappa > 0$ and a $\m Q_0 \succ 0$, then the system \eqref{eq:closed_loop_error_dyn} is exponentially stable with control law \eqref{e:input}. If \eqref{eq:step_2} is otherwise infeasible, then other possible values for $\kappa$ and $\m Q_0$ have to be sought and used until \eqref{eq:step_2} has at least a feasible solution. Unfortunately, finding a systematic method to find $\kappa$ and $\m Q_0$ such that \eqref{eq:step_2} is feasible---even though the existence of a solution can be guaranteed---is difficult and beyond the scope of the paper. To that end, this research direction is left for future research.  }\\
\indent Note that the solvability of the problem described by \eqref{eq:step_2} does not necessarily give a rapid closed-loop stabilization's convergence rate. Hence, to improve it, we propose the following approach to optimize the convergence rate, which in principle, solving \textbf{P1} by means of \textit{successive convex approximation} (SCA), which is pioneered in \cite{Dinh2012SCA}. The SCA basically approximates the BMIs with a series LMIs. In doing so, each of the BMI terms is firstly expressed as a difference between convex and concave functions. Later, the concave parts are over-approximated with their first order Taylor approximations.
This approach have been successfully implemented to tackle BMIs in several applications such as sensor and actuator selection problem \cite{Taha2019TCNS} and robust control of power networks \cite{Taha2019TCNS-2}.
To proceed with this approach, the nonconvex problem \textbf{P1} is transformed into an equivalent problem---provided below.
\begin{subequations}\label{eq:SDP_opt_prob_v2}
	\begin{align}
		(\mathbf{P2})\;\;&\minimize_{\m Q, \m K, \kappa, \epsilon,\alpha, t,w} \quad t \\
		&\subjectto \;\;\nonumber\\
		&\bmat{(\alpha-1)\m Q+\epsilon w\m I & * & *\\
			\m O & -\epsilon \m I &*\\
			\m A-\m B\m K & \m G & -\m Q^{-1}}\prec 0 \label{eq:SDP_opt_prob_v2_1}\\ 
		  &\bmat{-w & \gamma_x+\gamma_u\kappa \\ \gamma_x+\gamma_u\kappa &-1} \preceq 0,\;  \label{eq:SDP_opt_prob_v2_2} \\
		  &\eqref{eq:SDP_base_3},\,\eqref{eq:SDP_base_2},\, \m Q \preceq t \m I,\, t > 0, \, w\geq 0 \label{eq:SDP_opt_prob_v2_3}
	\end{align}
\end{subequations} 
The additional constraint \eqref{eq:SDP_opt_prob_v2_2} in \textbf{P2} above is a consequence of the over-bounding of the term  $\gamma_x+\gamma_u\kappa$ with a new scalar variable $w \geq 0$ such that $(\gamma_x+\gamma_u\kappa)^2 \leq w$. 

The problem \textbf{P2} is still highly nonconvex due to the terms $\m Q^{-1}$, $\alpha \m Q$ and  $\epsilon w$. To use the SCA, each of these terms need to be convexified. First, we decompose the bilinear term  $\alpha \m Q$ as follows
\begin{align}
\alpha \m Q = \dfrac{1}{4}\left(\left(\alpha \m I + \m Q\right)^\top \left(\alpha \m I + \m Q\right) - \left(\alpha \m I -\m Q\right)^\top \left(\alpha \m I - \m Q\right)\right). \label{eq:decomp_alpha_Q}
\end{align}
Define $\mathcal{H}(\alpha,\m Q) := - \left(\alpha \m I -\m Q\right)^\top \left(\alpha \m I - \m Q\right)$. Now, let $\tilde{\alpha}$ and $\tilde{\m Q}$ be the points of linearization. Since $\mathcal{H}(\cdot)$ is concave, then its first order Taylor approximation is a global over-estimator. As such, $\mathcal{H}(\cdot)$ can be over-approximated by a linear function
 $\mathcal{H}(\alpha,\m Q) \preceq \tilde{\mathcal{H}}(\alpha,\m Q,\tilde{\alpha},\tilde{\m Q})$ where 
 \begin{align*}
 	\begin{split}
\tilde{\mathcal{H}}(\alpha,\m Q,\tilde{\alpha},\tilde{\m Q}) &:= (\tilde{\alpha}^2-2\tilde{\alpha}\alpha)\m I + 2 \tilde{\alpha} \m Q + 2 \alpha \tilde{\m Q} \\
&\quad \tilde{\m Q}^2 - 2 \tilde{\alpha}\tilde{\m Q} - \tilde{\m Q}{\m Q} - {\m Q}\tilde{\m Q}.
 	\end{split}
 \end{align*}
From \eqref{eq:decomp_alpha_Q} and $\mathcal{H}(\alpha,\m Q) \preceq \tilde{\mathcal{H}}(\alpha,\m Q,\tilde{\alpha},\tilde{\m Q})$, we thus have
\begin{align}
\alpha \m Q \preceq \dfrac{1}{4}\left(\left(\alpha \m I + \m Q\right)^\top \left(\alpha \m I + \m Q\right) + \tilde{\mathcal{H}}(\alpha,\m Q,\tilde{\alpha},\tilde{\m Q})\right).\label{eq:decomp_alpha_Q_upper}
\end{align}
Second, we decompose the bilinear term  $\epsilon w$ as follows
\begin{align}
	\epsilon w = \dfrac{1}{4}\left((\epsilon+w)^2-(\epsilon-w)^2\right). \label{eq:decomp_eps_w}
\end{align}
Let $\mathcal{F}(\epsilon,w) := -(\epsilon-w)^2$. Its first order Taylor approximation can be represented by the following linear function
 \begin{align*}
	\hspace{-0.1cm}\begin{split}
		\tilde{\mathcal{F}}(\epsilon,w,\tilde{\epsilon},\tilde{w}) &:= \tilde{\epsilon}^2 + \tilde{w}^2 -2\tilde{\epsilon}\tilde{w}-2\epsilon(\tilde{\epsilon}-\tilde{w})-2w(\tilde{w}-\tilde{\epsilon}),
	\end{split}
\end{align*}
where $\tilde{\epsilon}$ and $\tilde{w}$ are the points of linearization. Since $\mathcal{F}(\epsilon,w) \leq \tilde{\mathcal{F}}(\epsilon,w,\tilde{\epsilon},\tilde{w})$ and from \eqref{eq:decomp_eps_w}, we get
\begin{align}
	\epsilon w \leq  \dfrac{1}{4}\left((\epsilon+w)^2+\tilde{\mathcal{F}}(\epsilon,w,\tilde{\epsilon},\tilde{w})\right). \label{eq:decomp_eps_w_upper}
\end{align}
Based on \eqref{eq:decomp_alpha_Q_upper} and \eqref{eq:decomp_eps_w_upper}, the left-hand side of \eqref{eq:SDP_opt_prob_v2_1} can be over approximated by the following matrix inequality
\begin{align}
\hspace{-0.3cm}\bmat{-\m Q+ \tfrac{1}{4}\left(\alpha \m I + \m Q\right)^\top \left(\alpha \m I + \m Q\right)\\ 
	+ \tfrac{1}{4}\tilde{\mathcal{H}}(\alpha,\m Q,\tilde{\alpha},\tilde{\m Q})  + \tfrac{1}{4}(\epsilon+w)^2\m I & * & *\\
	+\tfrac{1}{4}\tilde{\mathcal{F}}(\epsilon,w,\tilde{\epsilon},\tilde{w})\m I \\
	\m O & -\epsilon \m I &*\\
	\m A-\m B\m K & \m G & -\m Q^{-1}}\hspace{-0.1cm}\prec 0. \label{eq:SDP_main_upper}
\end{align}
Applying the Schur complement twice to \eqref{eq:SDP_main_upper}, we obtain
\begin{align}
	\hspace{-0.4cm}\bmat{-\m Q\\ 
		+ \tfrac{1}{4}\tilde{\mathcal{H}}(\alpha,\m Q,\tilde{\alpha},\tilde{\m Q}) & * & *&*&*\\
		+\tfrac{1}{4}\tilde{\mathcal{F}}(\epsilon,w,\tilde{\epsilon},\tilde{w})\m I \\
		\m O & -\epsilon \m I &*&*&*\\
		\m A-\m B\m K & \m G & -\m Q^{-1} &*&*\\
	\tfrac{1}{2}\left(\alpha \m I + \m Q\right) &\m O & \m O & -\m I  &* \\
\tfrac{1}{2}(\epsilon+w)^2\m I&\m O & \m O & \m O & -\m I }\hspace{-0.1cm}\prec 0. \label{eq:SDP_main_upper_schur}
\end{align}
Despite we have performed linearization towards the bilinear terms $\alpha \m Q$ and  $\epsilon w$, the matrix inequality is still nonconvex due to the term $\m Q^{-1}$. 
However, since $\m Q$ must be positive definite, then due to Lemma 1, we obtain
\begin{align}
	-\m Q^{-1}\preceq -2\tilde{\m Q}^{-1} + \tilde{\m Q}^{-1}\m Q\tilde{\m Q}^{-1}.\label{eq:Q_upper}
\end{align}
As such, from \eqref{eq:Q_upper} and \eqref{eq:SDP_main_upper_schur} and implementing a congruence transformation, one gets
 \begin{align}
	\hspace{-0.4cm}\bmat{-\m Q\\ 
	+ \tfrac{1}{4}\tilde{\mathcal{H}}(\alpha,\m Q,\tilde{\alpha},\tilde{\m Q}) & * & *&*&*\\
	+\tfrac{1}{4}\tilde{\mathcal{F}}(\epsilon,w,\tilde{\epsilon},\tilde{w})\m I \\
	\m O & -\epsilon \m I &*&*&*\\
	\tilde{\m Q}(\m A-\m B\m K) & \tilde{\m Q}\m G & -2\tilde{\m Q}+\m Q &*&*\\
	\tfrac{1}{2}\left(\alpha \m I + \m Q\right) &\m O & \m O & -\m I  &* \\
	\tfrac{1}{2}(\epsilon+w)^2\m I&\m O & \m O & \m O & -\m I }\hspace{-0.1cm}\prec 0, \label{eq:SDP_main_upper_LMI}
 \end{align}
which is an LMI. Given this result, our approach to solve \textbf{P2} my means of SCA is summarized in Algorithm \ref{algorithm1}. If a feasible initial solution is found and some assumptions hold, Algorithm \ref{algorithm1} is guaranteed to converge to a local optimal solution \cite{Dinh2012SCA}.   


\setlength{\floatsep}{12pt}
{
	\begin{algorithm}[t]
		\caption{Solving SCA for Controller Synthesis}\label{algorithm1}
		\DontPrintSemicolon
		\textbf{input:} $\m A$, $\m B$, $\m G$, $\gamma_x$, $\gamma_u$, $\mathrm{tol}$, and $\mathrm{MaxIter}$ \;
		\textbf{initial solution:} obtain a feasible solution to the problem described by \eqref{eq:step_2}; let $\m Q_0$, $\m K_0$, $\epsilon_0$, $\alpha_0$, $\kappa_0$ be the corresponding solutions \;
		\textbf{set:} $k := 1$, $t_0 = \lambda_{\mathrm{max}}(\m Q_0)$, and $w_0 := (\gamma_x + \gamma_u \kappa_0)^2+\varepsilon$ where $\varepsilon> 0$ is a relatively small constant  \;
		\While{$k \leq \mathrm{MaxIter}$}{
			\textbf{substitute:} $\tilde{\m Q} \leftarrow \m Q_{k-1}$, $\tilde{\alpha} \leftarrow \alpha_{k-1}$, $\tilde{\epsilon} \leftarrow \epsilon_{k-1}$, $\tilde{w} \leftarrow w_{k-1}$, $\tilde{t} \leftarrow t_{k-1}$  \;
			\textbf{solve:}
			\vspace{-0.1cm}
			\begin{subequations}\label{problem0} {
					\begin{flalign}
						(\mathbf{P3})\;&\minimize_{\m Q, \m K, \kappa, \epsilon,\alpha, t,w} \quad t \\
						&\subjectto \;\eqref{eq:SDP_main_upper_LMI},\, \eqref{eq:SDP_opt_prob_v2_2},\,
						\eqref{eq:SDP_opt_prob_v2_3}
				\end{flalign}}
			\end{subequations}\vspace{-0.0cm}
			\eIf{$|t-t_{k-1}| < \mathrm{tol}$}{
				break \;
			}{
				\textbf{substitute:} ${\m Q}_k \leftarrow \m Q$, ${\alpha}_k \leftarrow \alpha$, ${\epsilon}_k \leftarrow \epsilon$, ${w}_k \leftarrow w$, ${t}_k \leftarrow t$  \;
				\textbf{update:}	$k \leftarrow k + 1$\;
			}
		}
		\textbf{output:} $\m K$\;
\end{algorithm} \vspace{-0.2cm}}

\vspace{-0.2cm}
\section{Numerical Examples}\label{sec:simulationresults}
Herein we apply the designed controllers to two examples of nonlinear systems from the literature to show the effectiveness of the proposed approach. All convex problems with LMIs are solved by using YALMIP \cite{Lofberg2004} optimization interface through MATLAB along with MOSEK \cite{Andersen2000} solver.

\begin{figure}
	\centering
	\includegraphics[keepaspectratio=true,scale=0.685]{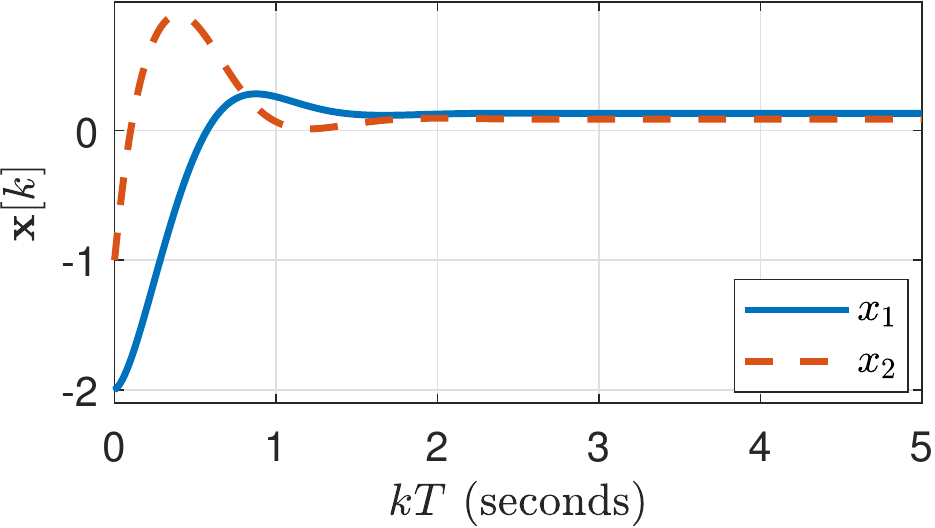}
	\caption{Stabilized trajectories of state variables for the system in Example 1 using controller without tracking. It can be seen that both states are converging to some nonzero equilibria.}
	\label{fig:Example1_Results_wotracking}
	\vspace{-0.1cm}
\end{figure}

\begin{figure}
	\centering
	\includegraphics[keepaspectratio=true,scale=0.685]{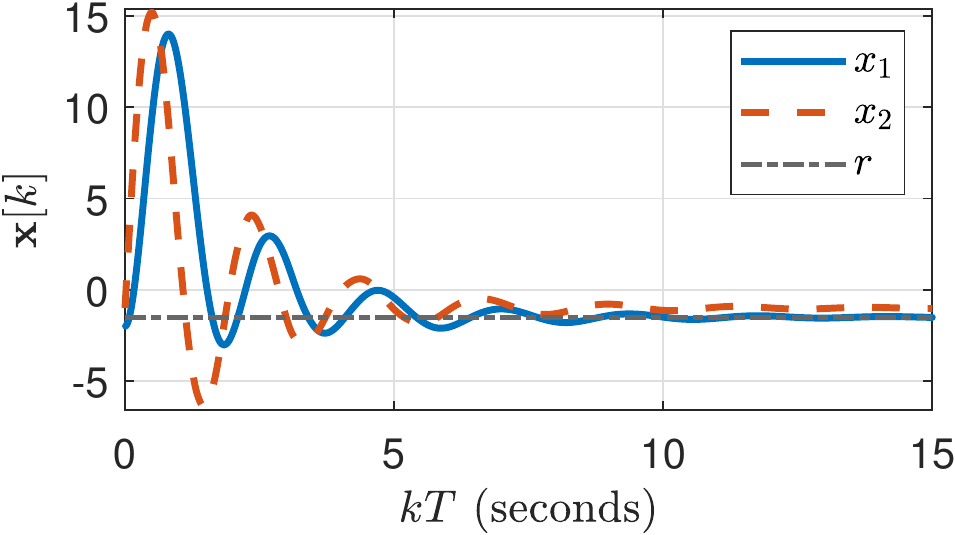}
	\caption{Stabilized trajectories of state variables for the system in Example 1 using controller with tracking for state $x_1$. Note that $x_1$ converges to the specified value of $r$ that is $-1.5$.}
	\label{fig:Example1_Results_tracking}	\vspace{-0.3cm}
\end{figure}

\vspace{-0.2cm}
\subsection{Example 1}\label{s:Example1}
In this example, we apply the proposed controllers to an unstable nonlinear system similar to \cite{PertewA.M2006HDfL,ekramian2017observer}. The system is originally defined in continuous-time which can be represented in the following state-space form
\begin{subequations}\label{e:ctsystem}
\begin{align}
    \dot{\m x}(t)&= \m A\m x(t)+\m f(\m x(t),u(t))+\m B u(t)\\
    \m y(t)&=\m C\m x(t),
    \end{align}
\end{subequations}

For this system, $\m x=\bmat{x_1 & x_2}$, that is there are two state variables, while there is a single input $ u$. The state-space parameter matrices for this system are given as
\begin{align*}
    \m A&=
    \begin{bmatrix}
    -2 &3\\
    3 &1\\
    \end{bmatrix},\hspace{2mm}
    \m B=[0 \hspace{2mm} 1]^\top, \hspace{2mm} \m C= 
    \begin{bmatrix}
    1&    0\\
    \end{bmatrix},\\
    \m f(\m x,u)&=
    \begin{bmatrix}
    0\\
    \gamma_x\cos(x_1-(\gamma_u/\gamma_x)u),\\
    \end{bmatrix},
\end{align*}
where $\gamma_x=1$ and $\gamma_u=0.1$ are the Lipschitz constants. We apply Euler discretization to obtain a discrete-time formulation for the continuous-time system given above, so that it may be represented by \eqref{e:system}. The obtained matrices are as follows
\begin{align*}
    \m A&=
    \begin{bmatrix}
    1-2T & 3T\\
    3T & 1+1T\\
    \end{bmatrix},\hspace{2mm}
    \m G= 
    \begin{bmatrix}
    T & 0\\
    0 & T\\
    \end{bmatrix},\\
    \m B&=[0 \hspace{2mm} T]^\top, \hspace{2mm} \m C= 
    \begin{bmatrix}
    1&    0\\
    \end{bmatrix},\\
    \m f(\m x[k],u[k])&=
    \begin{bmatrix}
    0\\
    \gamma_x\cos(x_1[k]-(\gamma_u/\gamma_x)u[k]),\\
    \end{bmatrix},
\end{align*}
where $T=0.01$ is chosen such that the states of the discrete-time system are as close to the continuous-time system as possible. The above system is highly unstable without any input.\\
\indent We first apply the proposed controller without output tracking to stabilize the state trajectories for this system. To obtain the controller gain $\m K$, we solve the problem $\mathbf{P2}$ using Algorithm \ref{algorithm1}. To get the initial solution, we first solve \eqref{eq:step_1} with $\alpha =10^{-2}$ and $\varrho =-20$ which gives us
\begin{align*}
    \m Q_0 =\bmat{0.0015 &   0.0009\\
    0.0009  &  0.0020}.
\end{align*}
Then, solving \eqref{eq:step_2} with $\kappa_0 = 10$ and $\m Q_0$ as above, we get $\alpha_0=1.47\times 10^{-2}$, $\epsilon_0=7.74\times 10^{-6}$ which completes our initial solution. Using $\varepsilon=0.01$, we compute $w_0=4.01$. We then run Algorithm \ref{algorithm1} starting with the above initial values to obtain the final value of gain $\m K$, which comes out as
\begin{align*}
    \m K=\bmat{-8.7744 &  -4.7690}.
\end{align*}
We use the computed gain $\m K$ to stabilize the state trajectories of the system which are presented in Figure \ref{fig:Example1_Results_wotracking} starting with an initial state $\m x[0] = \bmat{-2 &-1}$. 

Next, we apply the controller with output tracking to stabilize the system and make the state $x_1$ converge to an arbitrary reference value $r=-1.5$. To obtain the controller gain, we again solve $\mathbf{P2}$ but this time for an augmented system dynamics as shown in Section \ref{sec:controllerdesign}. The augmented state vector and state-space matrices for the given system, with $\m E=10^{-3} \m I$, are detailed as follows
\begin{align*}
    &\tilde{\m x} =\bmat{x_1 &x_2 &r},
    \hspace{5.5mm}\tilde{ \m A}=
   \begin{bmatrix}
    1-2T & 3T &0\\
    3T & 1+T &0\\
    10^{-3} &0 &1\\
    \end{bmatrix},\\
    &\tilde{\m G}= 
    \begin{bmatrix}
    T & 0 &0\\
    0 & T &0\\
    0 &0 &10^{-3}\\
    \end{bmatrix},
    \tilde{\m B}=\bmat{0 & T &0}^\top, \hspace{2mm}
    \tilde{\m C}= 
    \begin{bmatrix}
    0  &  1 &0\\
    \end{bmatrix},\\
    &\tilde{\m f}(\m x[k],u[k])=
    \begin{bmatrix}
    0\\
    \gamma_x\cos(x_1[k]-(\gamma_u/\gamma_x)u[k])\\
    -r\\
    \end{bmatrix}.
\end{align*}
We solve \eqref{eq:step_1} for this augmented system with $\alpha =  7\times 10^{-3}$ and $\varrho = -20$, which gives us
\begin{align*}
    \m Q_0 = 10^{-3}\times\bmat{0.0032  &  0.0009  &  0.0191\\
    0.0009  &  0.0026 &   0.0125\\
    0.0191   & 0.0125  &  0.1786}.
\end{align*}
Further, solving \eqref{eq:step_2} with $\kappa_0 =40 $ and $\m Q_0$ as above, we get $\alpha_0=3.1\times 10^{-3}$, $\epsilon_0=3.15\times 10^{-10}$, and using $\varepsilon=0.01$, we compute $w_0=25.01$. Running Algorithm \ref{algorithm1} starting with the above initial values, we get the final value of gain $\m K$, which comes out as
\begin{align*}
    \m K=\bmat{-7.3724 &  -3.6017 & -36.5141}.
\end{align*}
The stabilized state trajectories for this system using the output tracking controller are presented in Figure \ref{fig:Example1_Results_tracking} starting with the same initial state as before.


\vspace{-0.2cm}
\subsection{Example 2}\label{s:Example1}
\begin{figure}
    \centering
    \includegraphics[keepaspectratio=true,scale=0.685]{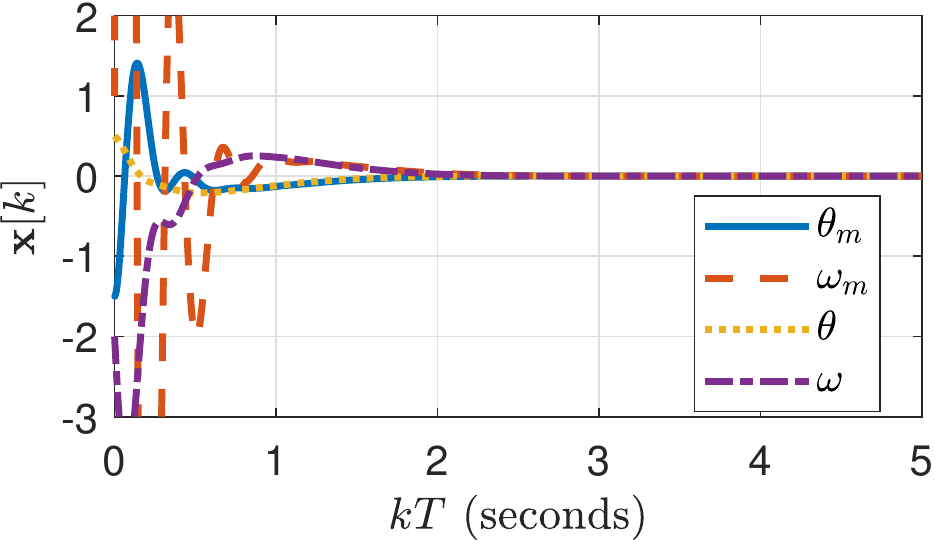}
    \caption{Stabilized trajectories of state variables for the system in Example 2 using controller without tracking. All states are converging to zero equilibria.}
    \label{fig:Example2_Results_wotracking}
\end{figure}

Herein we apply the proposed controller to a flexible link robot similar to \cite{PagillaP.R2004Caod,YadegarMeysam2018Otcd,ekramian2017observer}. Again the system is defined in continuous-time and has the form \eqref{e:ctsystem}. For this system we have 
\begin{align*}
    \m x&=\bmat{\theta_m & \omega_m & \theta & \omega},\\
    \m A&=
    \begin{bmatrix}
    0 &1 &0 &0\\
    -48.6 &-1.25 &48.6 &0\\
    0 &0 &0 &1\\
    1.95 &0 &-1.95 &0\\
    \end{bmatrix},\\
    \m B &= [0 \hspace{2mm} 21.6 \hspace{2mm} 0 \hspace{2mm} 0]^\top, \;\m C=[0 \hspace{2mm} 0 \hspace{2mm} 1 \hspace{2mm} 0],\\
    \m f(\m x)&=[0 \hspace{2mm} 0 \hspace{2mm} 0 \hspace{2mm} -0.25\sin(\theta)]^\top,
\end{align*}

where $\theta_m$ is the angular position of the motor, $\omega_m$ is the angular velocity of the motor, $\theta$ is the angular position of the link, and $\omega$ is the angular velocity of the link. The Lipschitz constants for this system are $\gamma_x=0.25$ and $\gamma_u=0$. The matrices for the discrete system \eqref{e:system} can be obtained similar to Example 1 (not shown here).
We set $T=0.001$ to make the states of the discrete-time system as close to the continuous-time system as possible. For the controller without output tracking, we start Algorithm \ref{algorithm1} with $\alpha = 10^{-3}$ and $\varrho = -5$ to obtain $\m Q_0$ (not presented here). Further, solving \eqref{eq:step_2} with $\kappa_0=1$ and obtained $\m Q_0$, we get $\alpha_0=1.83\times 10^{-4}$, $\epsilon_0=1.17\times 10^{-7}$, and using $\varepsilon=0.01$, we compute $w_0=0.0725$. Solving the iteration steps in Algorithm \ref{algorithm1} starting with the above initial values, we obtain the controller gain
\begin{align*}
    \m K=\bmat{-25.8500 &  -0.9142 &  16.7354  & -4.1012}.
\end{align*}
Applying the computed gain $\m K$ to the system, we stabilize the trajectories of the state variables starting from an initial state $\m x[0] = \bmat{-1.5 & 1 & 0.5 & -2}$ as shown in Figure \ref{fig:Example2_Results_wotracking}.
	\vspace{-0.0cm}

\begin{figure}
    \centering
    \includegraphics[keepaspectratio=true,scale=0.685]{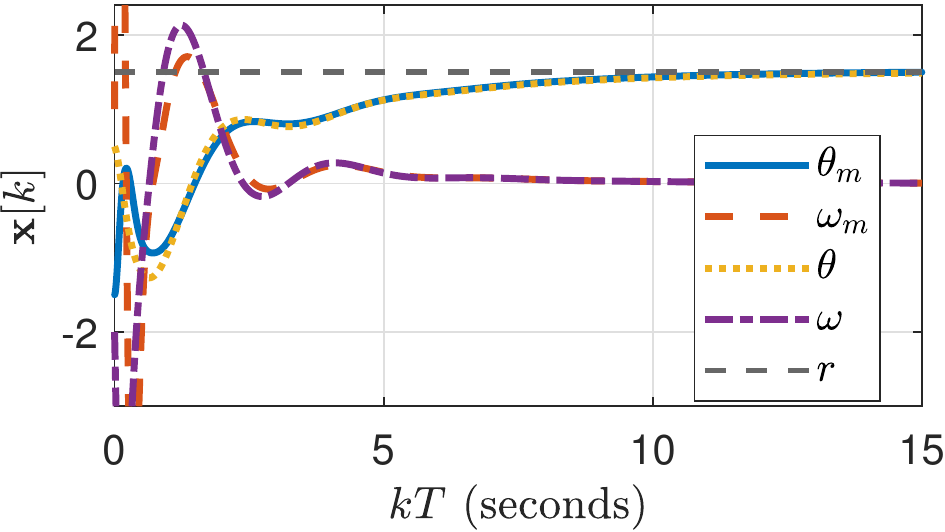}
    \caption{Stabilized trajectories of state variables for the system in Example 2 using controller with tracking for state $\theta$. Note that $\theta$ converges to the specified value of $r$ that is $1.5$.}
    \label{fig:Example2_Results_wtracking}	\vspace{-0.1cm}
\end{figure}

Next, we apply the controller with output tracking to stabilize the system and make the state $\theta$ converge to an arbitrary value $r=1.5$. The augmented state vector and state-space matrices for the given system can again be obtained similar to Example 1 (not shown here) with $\m E=10^{-3} \m I$.
 We solve \eqref{eq:step_1} for the augmented system with $\alpha=10^{-4}$ and $\varrho=-5$, which gives us the matrix $\m Q_0$ (not presented here). Further, solving \eqref{eq:step_2} with $\kappa_0=1$ and obtained $\m Q_0$, we get $\alpha_0=1.14\times 10^{-21}$, $\epsilon_0=1.56\times 10^{-24}$,
 and using $\varepsilon=0.01$, we compute $w_0=0.0725$. By running Algorithm \ref{algorithm1} starting with the above initial values we obtain the controller gain
\begin{align*}
\m K=\bmat{-6.9611 &  -0.7264  &  4.3826 &  -0.3681 &  -0.7463}. 
\end{align*}
The stabilized state trajectories for this system using the output tracking controller are presented in Figure \ref{fig:Example2_Results_wtracking} starting with the same initial state as before.

\section{Concluding Remarks and Future Work}\label{sec:conclusion}
In this paper we present a new sufficient condition to achieve exponential stability for discrete-time Lipschitz nonlinear systems. The prominent feature of our approach is its ability to accommodate nonlinear systems where the states and inputs are nonlinearly coupled. The proposed methodology can also be directly tailored for output regulation purpose. In addition to this, we present a simple algorithm to improve the convergence rate of the closed-loop performance which only requires the utilization of a convex programming solver. Potential future research directions include extending the proposed approach for nonlinear discrete-time systems which nonlinearity belongs to function sets other than Lipschitz and developing a robust stabilization method to tackle disturbance and unknown inputs.

\vspace{-0.05cm}

\bibliographystyle{IEEEtran}	\bibliography{bibl}

\end{document}